# Electrically pumped polarized exciton-polaritons in a halide perovskite microcavity

Tingting Wang[1,2,†], Zhihao Zang[1,2,†], Yuchen Gao[1], Chao Lyu[1], Kai Peng[3], Kenji Watanabe[4], Takashi Taniguchi[5], Xiaoze Liu[6], Wei Bao[3,*], and Yu Ye[1,2,7*]

[1]State Key Laboratory for Mesoscopic Physics and Frontiers Science Centre for Nano-optoelectronics, School of Physics, Peking University, Beijing 100871, China
[2]Collaborative Innovation Centre of Quantum Matter, Beijing 100871, China
[3]Electrical & Computer Engineering, University of Nebraska-Lincoln, Nebraska 68588, USA
[4]Research Center for Functional Materials, National Institute for Materials Science, 1-1 Namiki, Tsukuba, 305-0044, Japan
[5]International Center for Materials Nanoarchitectonics, National Institute for Materials Science, 1-1 Namiki, Tsukuba, 305-0044, Japan
[6]School of Physics and Technology, Wuhan University, Wuhan 430072, Hubei, China
[7]Peking University Yangtze Delta Institute of Optoelectronics, Nantong 226010, Jiangsu, China
[†]These authors contributed equally to this work
[*]Correspondence and request for materials should be addressed to W.B. (wbao@unl.edu) and Y. Y. (ye_yu@pku.edu.cn)

**Abstract**

Exciton polaritons, hybrid quasiparticles with part-light part-matter nature in semiconductor microcavities, are extensively investigated for striking phenomena such as polariton condensation and quantum emulation. These phenomena have recently been discovered in emerging lead halide perovskites at elevated temperatures up to room temperature. For advancing these discoveries into practical applications, one critical requirement is the realization of electrically pumped exciton-polaritons. However, electrically pumped polariton light-emitting devices with perovskites have not yet been achieved experimentally. Here, we devise a new method to combine the device with the microcavity and report the first halide perovskite polariton light-emitting device. Specifically, the device is based on a $CsPbBr_3$ capacitive structure, which can inject the electrons and holes from the same electrode, conducive to the formation of excitons, maintaining the high quality of the microcavity. In addition, highly polarization-selective polariton emissions have been demonstrated due to the optical birefringence in $CsPbBr_3$ microplate. This work paves the way for realizing practical polaritonic devices such as high-speed light-emitting devices for information communications and inversionless electrically pumped lasers based on perovskites.

**Introduction**

Lead halide perovskites provide an ideal platform for optoelectronic owing to their outstanding physical properties, including considerable exciton binding energy at room temperature, large oscillator strength, high carrier mobility, low nonradiative recombination rate, tunable bandgap, long carrier lifetime, high optical gain, etc[1–8]. Along with the fast-paced developments of photovoltaics and lighting applications, electrically pumped lasers have been a long-sought goal in the research field of perovskite optoelectronics. However, the lasing action usually requires a high enough density of excited carriers for the population inversion of stimulated emission, which has been extremely challenging for the electrical pumping in perovskites so far. This high injection requirement could be significantly relaxed (~ two orders of magnitude smaller) if the lasing action results from the stimulated scattering of exciton-polaritons[9]. However, the first step, the electrically-pumped exciton-polariton of the perovskite materials, has not yet been realized.

Exciton-polaritons are formed under the strong-coupling regime in semiconductor cavities, where the coupling rate between exciton and photon is much faster than the average decay rates of excitons and cavity photons. The hybridization makes the exciton-polaritons possess a much small effective mass and strong nonlinear interaction strength, leading to the discoveries of fascinating physics such as polariton condensation and quantum emulation at elevated temperatures compared to the ultra-cold atoms[10]. These physical phenomena have been subsequently discovered in perovskite microcavities, resulting in its renaissance in the past few years[1,11–24], e.g. polariton lasing and condensation at room temperature[11,22], the Rydberg exciton-polaritons (REPs)[23,25], and polariton lattices for quantum emulation and topological physics[1,18–21]. All the progress shows the great potential of perovskite polaritonics in fundamental research and practical applications. However, no works have shown that the perovskite exciton-polaritons could be electrically pumped so far even though the solution-processed perovskite light-emitting diodes (LEDs) have recently achieved big strides[26–29]. This is because the realization of polariton light-emitting devices based on solution-processed perovskite materials present some challenges compared to two-dimensional semiconductors[30] or semiconductor multiple quantum wells[31], including i) The solution-processed polycrystalline perovskite cannot reliably maintain a exciton state, which hinders the formation of exciton-polaritons; ii) The thickness of each layer in spin-coated perovskite LEDs cannot be precisely controlled to form a single-mode emission, which is essential for achieving exciton-polariton condensation; iii) The current device structure and fabrication process cannot guarantee a smooth interface coupled with a high-quality microcavity.

To address these challenges, we realize the electrically pumped polariton light-emitting device by assembling inorganic perovskite $CsPbBr_3$ microplates with hexagonal boron nitride (*h*-BN) and few-layer graphite (FLG) in a capacitive structure. Thus, by applying an alternating current (AC) voltage, the electrons and holes can be separately injected into the $CsPbBr_3$ microplates from the FLG alternatively in time but close in space. To form a high-quality microcavity, we develop a new approach to deterministically transfer the silver top mirror. The key feature of this technique is that the materials' interfaces do not contact any organic polymers and solvents during the transfer process, thereby avoiding the introduction of residues in the microcavity and ensuring the smoothness of the interfaces. The transferred silver fragments can be used not only as a mirror[23,24,30,32] but also as an electrode, which can be generalized as a universal technique to meet the needs of any size and position of silver fragments in nano-optoelectronic devices. In this way, the control of the cavity layer's thickness could also be taken care of for the cavity resonance condition. The realization of an electrically pumped polariton light-emitting device based on $CsPbBr_3$ microplate presents a critical step toward ultrafast polariton light emission for communications and electrically pumped inversionless polariton laser based on perovskites.

## Results

To construct a polariton light-emitting device, we assembled a hybrid structure of *h*-BN/FLG/$CsPbBr_3$/*h*-BN on the DBR substrate by using an all-dry transfer method[33,34]. Then we used the developed deterministic dry-transfer method (see more details in Methods) to assemble a micron-scale silver flake onto the hybrid structure (Fig. 1a), acting as a top mirror and an electrode of the light-emitting device at the same time. The bottom DBR mirror is composed of 14.5 pairs of $Ta_2O_5/SiO_2$ structures, showing reflectivity of over 99.94%, and a stopband ranging from 450 nm to 615 nm (see Fig. S1a). Meanwhile, the transferred silver mirror also shows adequate reflectivity. The bare microcavity composed of the bottom DBR and the transferred silver mirror with a 220 nm thick polymethyl methacrylate (PMMA) spacer layer shows a resonance at 537 nm with a full width at half maximum (FWHM) of 0.89 nm and a passive cavity quality factor (*Q* factor) of ~ 603 (see Fig. S1b). In the polariton light-emitting device, the active layer of the $CsPbBr_3$ microplate and the dielectric layers of *h*-BN are carefully selected, and their thickness is accurately measured by atomic force microscopy (AFM) to reach a $5\lambda/2$ cavity mode that resonates with the $CsPbBr_3$ exciton energy (see Methods and Supplementary Section 2 in detail). The typical device's optical microscope image (top view) is shown in Fig. 1c, with the schematic illustration of the device's structure in Fig. 1b. To avert uncoupled exciton emission and AC capacitive leakage current, we precisely place the top silver flake (marked by the grey dashed line) to completely cover the CsPbBr$_3$ microplate, while minimizing its overlap with the bottom FLG electrode (marked by the black dashed line in Fig. 1c). Due to its orthorhombic crystal structure[35], CsPbBr$_3$ has a mutually orthogonal birefringence effect along with the *a* and *b* crystalline axes, resulting in two polarization modes when coupled to microcavities[18,20,23].

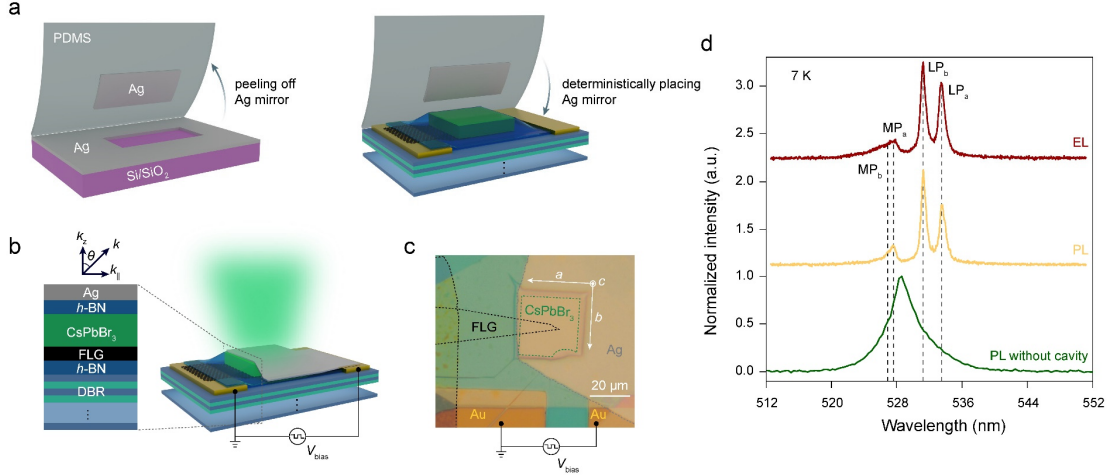

**Fig. 1 Device schematic and characterization. a**, Schematic of device assembly processes. The Ag mirror is first peeled off from the Ag film previously deposited on a Si/SiO$_2$ substrate and then deterministically placed on the fabricated device using a dry-transfer method. The specific stacking process is given in Supplementary Section 2. **b**, Schematic illustration of the final device's structure with the cross-sectional view shown in the inset. **c**, Optical microscope image of a typical device. **d**, EL (brown trace) and PL (yellow trace) spectra of the CsPbBr$_3$ polariton light-emitting device recorded at 7 K and $\theta = 0°$. PL (Olive green trace) spectrum of the CsPbBr$_3$ microplate on DBR substrate is also provided for comparison. The applied voltage $V_{bias}$ is 20 V with a frequency ($f$) of 1 MHz for EL measurement, while the optical pump power is 10 μW for PL measurement. The vertical black dashed lines indicate the main emission modes MP$_b$, MP$_a$, LP$_b$, and LP$_a$.

For the EL measurements, we applied a bipolar square wave AC voltage ($V_{bias}$) greater than the turn-on voltage to the top electrode (silver flake), while the bottom electrode (FLG) was grounded (Fig. 1b and 1c). Due to the long carrier lifetime and the large exciton binding energy of CsPbBr$_3$ at low temperatures, the alternately injected electrons and holes will diffuse in the microplate and form excitons. In a recent experimental work[23], REPs (corresponding exciton component of the 1s and 2s states) were observed in a single-crystal CsPbBr$_3$ microcavity without an external field. When more than one excitonic states couple to the same optical mode, hybrid polariton states appear. In order to minimize the heating effect caused by the external field, the photoluminescence (PL) and EL spectra of the device were recorded at 7 K

(all the temperature below is 7 K except for special indication) and $\theta = 0°$ ($\theta$ refers to the angle of collected emission direction with relative to the z-axis) are consistent with each other, and four emission peaks can be clearly seen (Fig. 1d), distinct from the PL emission of uncoupled CsPbBr$_3$ microplate. The featureless PL of a CsPbBr$_3$ microplate placed on the DBR substrate is also provided for comparison (Fig. 1d). In a strong-coupling hybrid polariton system, these four peaks can be well-referred to as the lower polariton branches (LP$_a$ and LP$_b$) and the middle polariton branches (MP$_a$ and MP$_b$), where the subscripts, a and b, are attributed to the orthogonal birefringence along the two crystal axes, *a*-axis and *b*-axis. For the electrically pumped lower polariton branches, LP$_a$ (LP$_b$) is in well resonant with the cavity, with a center at 533.5 nm (531.2 nm), an FWHM of 0.89 nm (0.71 nm), and a *Q* factor of ~ 601 (745). However, due to their large Rabi splitting, strong absorption above the bandgap, the upper polariton branches (UP$_a$ and UP$_b$) cannot be observed here, which is also very common in other inorganic perovskite microcavities[11,23].

For further confirmation of the light-matter coupling and illustration of the polariton emission of the CsPbBr$_3$ light-emitting device, we used home-built k-space spectroscopy (see Supplementary Section 3) to map out the dispersion relationship between wavelength (energy) versus angle (in-plane momentum). In the angle-resolved reflectivity map (Fig. 2a), we can clearly distinguish the middle and lower polariton branches, while the upper branches are barely seen. The dispersion curvatures of the MP and LP branches are unambiguously flattened at larger angles, and their resonance frequency shows anti-crossing behaviors from parabolic FP cavity dispersion and flat perovskite exciton energy, firmly indicating that the strong-coupling regime has been reached. Since the focused laser spot reduces the inhomogeneity of the emission, the angle-resolved PL map excited under the non-resonant excitation of the 405 nm continuous-wave (CW) laser (left panel in Fig. 2b) shows more pronounced strong-coupling features compared to those from the reflectivity map. There is a tiny energy difference between these two spectra, which can be attributed to the inhomogeneity of the structure. The emission intensity with wavelength larger than the CsPbBr$_3$ 1s exciton emission (~ 529.4 nm) in the PL and EL maps is amplified by a factor of five to enhance the visibility of the middle polariton branches (MP$_b$ and MP$_a$). The dispersion curves of the polariton branches can only be well fitted by the three-coupled oscillator model[23] (more details on theoretical fitting are given in Supplementary Section 4), that is, two exciton states are coupled to one optical mode (Fig. 2b). These two exciton states have been reported in the previous optical pumping case[23]. As the temperature increases, the middle polariton branches gradually become blurred and eventually disappear above 150 K (see Supplementary Section 5), and the hybrid two-exciton polariton state becomes a single-exciton polariton state because the thermal energy dissociates the 2s excitons with smaller exciton binding energy. By applying a bipolar square voltage with a $V_{bias}$ of 20 V and an $f$ of 1 MHz, we obtain strong EL. The angle-resolved EL map (right panel in Fig. 2b) matches with the PL map well, which firmly illustrates that the electrically pumped strong-coupling regime has been reached, and the pulsed electrical injection does not cause significant Joule heating. Using the three-coupled oscillator model, we can obtain the cavity detuning ($\Delta = E_c - E_x$, where $E_c$ is the cavity photon energy at zero in-plane momentum and $E_x$ is the exciton energy) between the 1s (2s) exciton state and cavity mode along the $a$-axis and $b$-axis, respectively, that is, $\Delta_{1a} = E_{ca} - E_{1a} = 28.2$ meV ($\Delta_{2a} = E_{ca} - E_{2a} = -2.7$ meV) and $\Delta_{1b} = E_{cb} - E_{1b} = 11.2$ meV ($\Delta_{2b} = E_{cb} - E_{2b} = -1.8$ meV). In addition, we also obtain the Rabi splitting energy, $\hbar\Omega$, between the LP (MP) branch and the UP branch along the $a$-axis and $b$-axis, which are 37.4 meV and 24 meV (33 meV and 10.2 meV), respectively. These splitting energies are much larger than the corresponding polariton linewidth, which further justifies that our device is working in the strong-coupling regime. Based on the three-coupled oscillator model, the Hopfield coefficients of each polariton branch can be obtained, which provides the weight of each constituent (see Supplementary Section 6). Specifically, as the angle increases, the contribution of the excitonic components (i.e., 1s and 2s exciton states) in the LP and MP branches gradually increases, while the photonic components in the UP branches increases, consistent with the previous report[23].

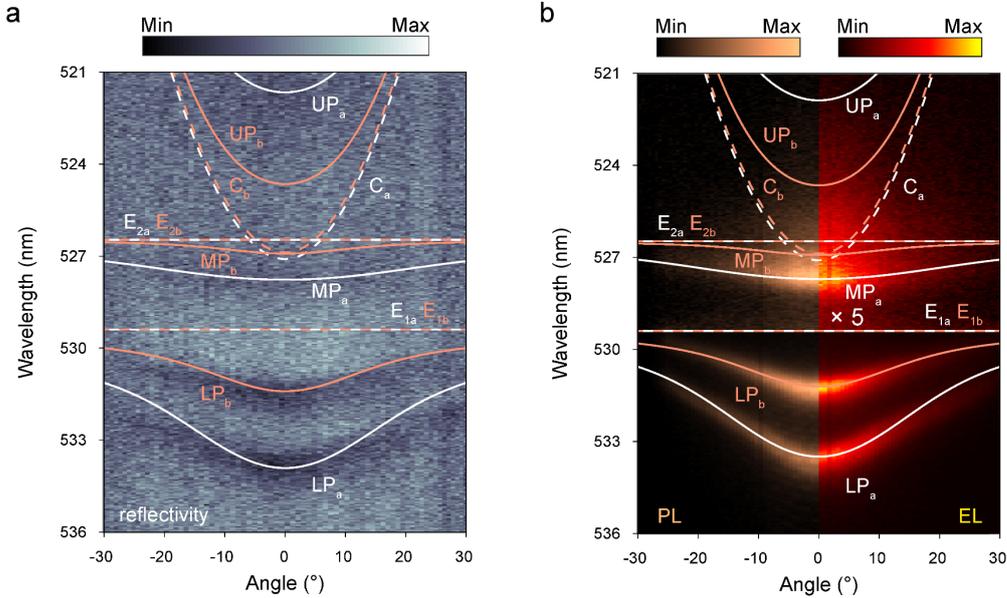

**Fig. 2 Characterizations of polariton emission dispersions. a, b,** Angle-resolved reflectivity (**a**), PL (left panel in **b**), and EL (right panel in **b**) map of the device measured at 7 K. The horizontal axis represents the angle of the emission relative to the $z$-axis, and the vertical axis is the wavelength. The white and orange solid lines display the theoretical fittings of the upper ($UP_a$ and $UP_b$), middle ($MP_a$ and $MP_b$),

and lower (LP$_a$ and LP$_b$) polariton dispersions along the *a*-axis and *b*-axis, respectively. The white and orange dashed lines depict the dispersions of uncoupled excitons (E$_{2a}$, E$_{2b}$, E$_{1a}$, and E$_{1b}$) and cavity photon mode (C$_a$ and C$_b$) obtained from a three-coupled oscillator model fitting along the *a*-axis and *b*-axis, respectively. For clarity, the intensity of the middle polariton branches (MP$_b$ and MP$_a$) in **Fig. 2b** is magnified by a factor of five. The angle-resolved reflectivity and PL maps match with the EL map well.

Due to the distinct birefringence effect in CsPbBr$_3$ along the *a*-axis and *b*-axis and the single-crystal microplate employed in our devices, it is expected that linearly polarized light sources will be implemented in our light-emitting device. To perform polarization-dependent angle-resolved EL measurements, we sequentially placed a half-wave plate (HWP) and a linear polarizer in the collection light path, and the polarization-dependent signal is collected by rotating the HWP. Figures 3a-c exhibit the angle-resolved EL at three quintessential polarization angles (along the perovskite *a*-axis, 45° between *a*-axis and *b*-axis, and *b*-axis), superimposed with the theoretical dispersion fittings. When the detection polarization is set along the *a*- (*b*-) axis, only the LP$_a$ (LP$_b$) and MP$_a$ (MP$_b$) can be observed, as shown in Fig. 3a (Fig. 3c). When the detection polarization is set along 45° between the *a*-axis and *b*-axis, both two sets of branches can be observed and linearly superimposed (see Fig. 3b), indicating they are mutually orthogonal to each other. Optically pumped linearly polarized polaritons in perovskites have previously been reported[23] (see the optically pumped results in our device in the Supplementary section 7), and recently polarization has been used as a new degree of freedom to switch between different topological polariton phases[18]. Now, it is realized by electrical means, which not only brings about the linearly polarized light source but also opens up a way to control the polaritons and their polarizations in perovskites electrically.

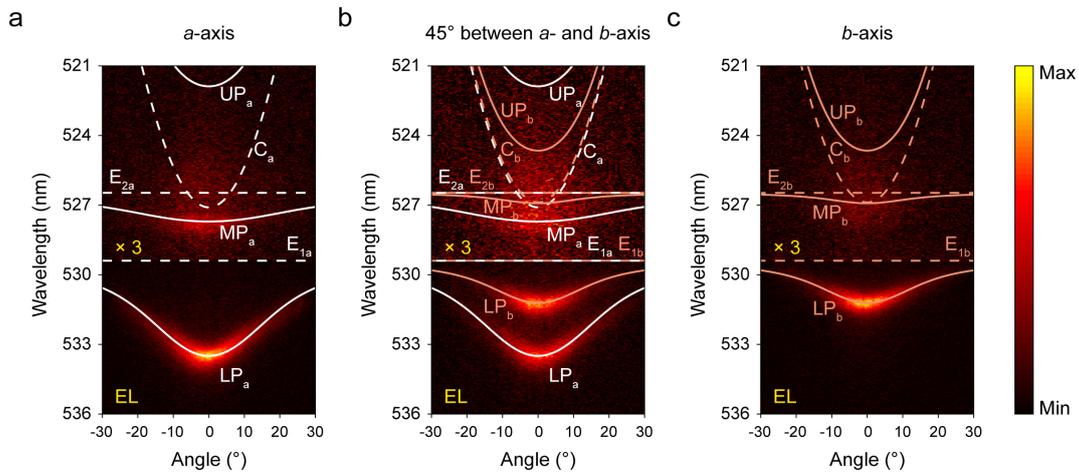

**Fig. 3 Polarization-resolved EL. a, b, c,** Angle-resolved EL maps under a $V_{\text{bias}}$ of 20 V with an *f* of 1 MHz, corresponding to the polarization direction along the *a*-axis (**a**),

45° between *a*-axis and *b*-axis (**b**), and *b*-axis (**c**), respectively. For better visibility, the intensity of the middle polariton branches is multiplied by three. Similar to **Fig. 2**, the white lines depict the polarization mode along the *a*-axis, while the orange lines depict the orthogonal polarization mode along the *b*-axis.

In our device structure, the pulse voltage can allow a very high injection current density, and the silver flake simultaneously as a mirror and an injection electrode so that the injection current will not affect the optical feedback structure. Therefore, we probed the dependence of the polariton EL intensity on the excitation voltage and frequency (Fig. 4). When the applied voltage frequency, *f*, is 1 MHz, the EL intensities of all four polariton branches ($MP_b$, $MP_a$, $LP_b$, and $LP_a$) exhibit a threshold behavior with the applied voltage (Fig. 4a). The turn-on voltage is about 13 V, and the EL intensity increases exponentially with the applied voltage after it is turned on. The obvious exponential correlation can be attributed to the increase in the number of injected carriers by the tunneling effect during the bipolar switching (see the band diagrams of the device under applied voltage in different states during a period in Supplementary Section 8)[36]. When the applied bipolar voltage $V_{bias}$ is constant at 18 V, the EL intensities of all four polariton branches increase linearly with the frequency (Fig. 4b). This is because the carrier injection occurs only during the bipolar switching (see Fig. S7), the EL intensity increases linearly with the number of voltage cycles, namely, the frequency[36]. To avoid the ion migration and exciton dissociation in perovskite under a large applied electric field, the applied voltage is kept at a low voltage level. It is worth noting that no saturation of EL intensity is observed here, indicating that our structure can realize a strong electrically pumped exciton-polariton polarized light source in perovskite. For further promoting the realization of lasing operation, the structure optimization of the light-emitting layer, the improvement of the material performance, the improvement of the microcavity's *Q* factor, and the reduction of the turn-on voltage (by tuning the thickness of the *h*-BN tunneling layer between the perovskite and the silver electrode), etc., are required.

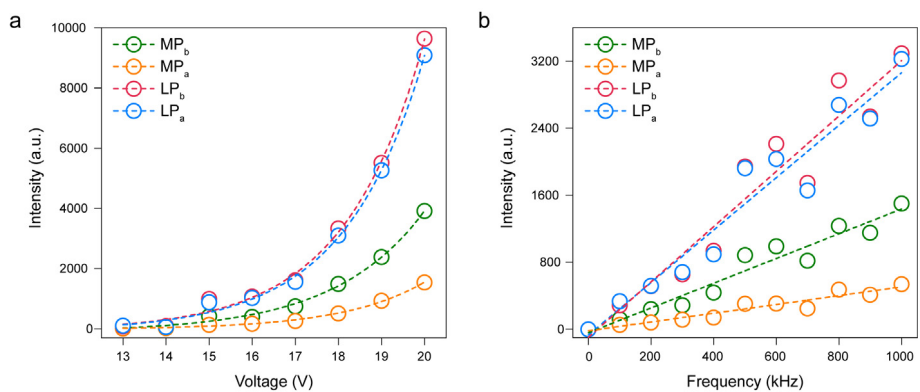

**Fig. 4 Voltage and frequency dependence of the polariton EL. a,** Voltage-dependent polariton integrated EL intensity under a constant *f* of 1 MHz. The dashed

curves depict the exponential fitting relationship of their respective integrated intensity with voltage. **b,** Frequency-dependent polariton integrated EL intensity under a constant $V_{bias}$ of 18 V. The relationship between integrated intensity and frequency is fitted by a linear function. The acquisition time is 2 s (**Fig. 4a** and **4b**).

**Conclusion**

In summary, we have demonstrated an electrically-driven $CsPbBr_3$ polariton light-emitting device based on a capacitive structure. A new approach of deterministic dry transfer of the silver flake has also been developed, which could be used as the thin top optical mirror and electrode simultaneously. By fine-tuning the thickness of the cavity, excitons and cavity photons can be strongly coupled to form bright electrically driven polariton emissions. Hybrid polariton states have been observed in our device. Due to the existence of two polarization modes induced by the birefringence effect of the employed microplate when coupled to a microcavity, linearly polarized polariton light sources were realized by electrical means. The powerful technology-material-device combination opens the door to perovskite-based polariton light sources and pushes the electrically pumped polariton perovskite lasers a step closer to reality.

**Methods**
**Device fabrication.**
The light-emitting device consists of a bottom DBR, the electrically-driven active region (composed of FLG, thin *h*-BN flakes, and a $CsPbBr_3$ microplate), and the transferred silver film. The bottom DBR is composed of 14.5 pairs of $SiO_2/Ta_2O_5$ deposited on silicon substrates via ion beam sputtering. The FLG and thin *h*-BN flakes on PDMS substrates are obtained by mechanical exfoliation from bulk crystals and transferred onto the bottom DBR via a dry-transfer method. The $CsPbBr_3$ microplates are synthesized on the mica substrates as described in our previous work[36] and transferred by the PDMS. The span-new method of deterministic dry transfer of silver film is developed. Before the silver fragments are mechanically peeled off and transferred onto the desired position by PDMS, the 30-nm silver film is evaporated on the Si/$SiO_2$ substrates using electron beam evaporation (EBE) with a deposition rate of 1 Å/s. The thicknesses of each layer in our device are confirmed by AFM (Cypher, Asylum Research) with a tapping mode. Cr (5 nm)/Au (30 nm) electrodes are deposited on the DBR substrates via EBE in advance for electrical measurements.

**Optical spectroscopy measurements.**

A schematic diagram of the home-built angle-resolved reflectivity, PL, and EL set-up is shown in Supplementary Section 3. The device is kept in a liquid Helium cryostat (Janis ST-500) for low-temperature measurements. A broadband white light source (Thorlabs SLS401) is used for angle-resolved reflectivity measurements as well as device imaging. In terms of angle-resolved PL measurements, the $CsPbBr_3$ light-emitting device is pumped by a CW laser (405 nm). The excitation light (laser) beam is focused by an objective (Zeiss Epiplan-Neofluar 50×/0.55 DIC M27). The emitted signals are collected by the same objective and sent into the Andor spectrometer (SR-500i-D2-R) equipped with gratings of 150- and 1200-lines $mm^{-1}$ and a Newton CCD (DU920P-BEX2-DD) of 256×1024 pixels.

**Electrical spectroscopy measurements.**
Our setup in the Fourier imaging configuration allows us to access the angle-resolved EL spectra with the help of a pulse generator (Agilent 8114A) providing the AC gate voltage.

**Author contributions**
Y.Y. conceived the project. T.W. and Z.Z. fabricated the devices and performed the measurements. Y.G. carried out the theoretical fittings. K.P. and W.B. grew the bottom DBR mirror. C.L. helped in metal electrode fabrication and wire bonding process. K.W. and T.T. grew the *h*-BN single crystals. T.W., Z.Z., and Y.Y. performed data analysis and wrote the paper with inputs from all authors. All authors discussed the results, data analysis, and the paper.

**Acknowledgements**
This work was supported by National Natural Science Foundation of China (No. 61875001) and the National Key R&D Program of China (grants 2018YFA0306900 and 2017YFA0206301). K.W. and T.T. acknowledge support from the Elemental Strategy Initiative conducted by the MEXT, Japan (Grant No. JPMXP0112101001), JSPS KAKENHI (Grant Nos. 19H05790, 20H00354, and 21H05233) and A3 Foresight by JSPS.

**Competing interests**
The authors declare no competing interests.

**Additional information**
Supplementary information is available for this paper at

**Data availability**

The data that support the findings of this study are available from the corresponding author upon request.